\documentstyle[aps,preprint,amsmath,amssymb]{revtex}
\begin{document}
\draft
%%%%%%%%%%%%%%%%%%%%%%%%%%%%%%%%%%%%%%%%%%%%%%%%%%%%%%%%%%%%%%%%%%%%
\title{Back Reaction Problem in the Inflationary Universe}
\date{September 27, 2000}
\author{Yasusada Nambu\footnote{e-mail:~nambu@allegro.phys.nagoya-u.ac.jp}}
\address{Department of Physics, Graduate School of Science, Nagoya University \\ 
        Chikusa, Nagoya 464-8602, Japan}
%
%%%%%%%%%%%%%%%%%%%%%%%%%%%%%%%%%%%%%%%%%%%%%%%%%%%%%%%%%%
\tightenlines
%%%% DPNU number %%%%
\preprint{DPNU-00-27}
%%%%%%%%%%
\maketitle
%%%%%%%%%%%%%%%%%%%%%%%%%%%%%%%%%%%%%%%%%%%%%%%%%%%%%%%%%%
\begin{abstract}
  We investigate the back reaction of cosmological perturbations on
  an inflationary universe using the renormalization-group
  method. The second-order zero mode solution which appears by
  the nonlinearity of the Einstein equation is regarded as  secular terms
  of a perturbative expansion, we redefine constants of
  integration contained in the background solution and absorbed the
  secular terms to these constants in a gauge-invariant manner. The
  resulting renormalization-group equation describes the back reaction
  effect of inhomogeneity on the background universe. For scalar type
  classical perturbation, by solving the renormalization-group
  equation, we find that the back reaction of the long wavelength
  fluctuation works as a positive spatial curvature, and the short
  wavelength fluctuation works as a radiation fluid. For the long
  wavelength quantum fluctuation, the effect of back reaction is
  equivalent to a negative spatial curvature.
\end{abstract}
%%%%%%%%%% local macro %%%%%%%%%%%
\def\al{\alpha}
\def\del{\delta}
\def\ep{\epsilon}
\def\pa{\partial}
\def\bk{\boldsymbol{k}}
\def\bx{\boldsymbol{x}}
%%%%%%%%%%%%%%%%%%%%%%%%%%%%%%%%%%%%%%%%%%%%%%%%%%%%%%%%%%%%%%
\section{introduction}
In the context of standard cosmological perturbation
approach\cite{kodama,mukhanov2}, our Universe is treated as a
homogeneous isotropic Friedmann-Robertson-Walker(FRW) model plus small
fluctuations on it.  But due to the nonlinearity of Einstein's
equation, the fluctuation has an effect on the evolution of the
background spacetime and we must solve the evolution of the
fluctuation and the background in a self-consistent manner. This is
the cosmological back reaction problem and has been studied by several
authors\cite{issacson,futamase1,futamase2,russ2,boersma,mukuhanov1,abramo1,abramo2,abramo3}.
The conventional approach to the problem is to construct an effective
energy momentum tensor of the fluctuation. By adding this tensor to
the right-hand side of the background Einstein equation, we can
evaluate the effect of inhomogeneity on the evolution of the
background FRW universe. For large scale fluctuation of which wave
length is larger than the Hubble horizon, gauge dependence of the
perturbation becomes conspicuous and it is necessary to construct a
gauge-invariant formalism of the back reaction problem. In this
direction, Abramo and
co-workers\cite{mukuhanov1,abramo1,abramo2,abramo3} derived the
gauge-invariant effective energy momentum tensor of cosmological
perturbations, which is invariant under the first-order gauge
transformation. They applied their formalism to an inflationary
universe and obtained the result that the back reaction effect of the
long wavelength scalar type fluctuation is equivalent to a negative
cosmological constant, and greatly reduces the inflationary expansion
of the universe. But they did not derive solutions of an effective
scale factor for the FRW universe with the back reaction.
Recently, the renormalization-group
method\cite{chen,kunihiro,nozaki,nambu} is applied to the cosmological
back reaction problem\cite{nambu2}. Starting from a naive perturbative
expansion of the solution of the original differential equation, this
method gives an improved solution by renormalizing a secular term
which appears by the nonlinearity of the equation. For a dust
dominated universe, the second-order gauge-invariant zero mode metric
is constructed by using the method of Abramo and co-workers. Then, by
assuming that the second-order metric is secular, it is absorbed in a
constant of integration contained in the background scale factor by
using the renormalization-group method. The renormalized scale factor
represents the effective dynamics of the FRW universe with the back
reaction due to inhomogeneity. By solving the renormalization-group
equation, it was found that perturbations of the scalar mode and the
long wavelength tensor mode work as a positive spatial curvature, and
the short wavelength tensor mode as a radiation fluid.

In this paper, we aim to investigate the back reaction problem in the
inflationary universe using the renormalization-group method. The
advantage of this method is that we can obtain a solution of the back
reaction equation directly by solving the renormalization-group
equation. We do not need to solve the FRW equation with the effective
energy momentum tensor, which is done in a conventional approach to
the back reaction problem.

The plan of this paper is as follows. In Sec. II, we introduce the
renormalization-group method by using a FRW model with a cosmological constant
and  perfect fluid. In Sec. III, the formulation of the back reaction
problem based on the renormalization group-method is presented. In Sec. IV, we
apply our formalism to the inflationary universe. Section V is devoted to a 
summary and discussion. We use the units in which $c=8\pi G=1$
throughout the paper. 
%%%%%%%%%%%%%%%%%%%%%%%%%%%%%%%%%%%%%%%%%%%%%%%%%%%%%%%%%%%%%%
\section{renormalization group method}
To introduce the renormalization-group method, we consider a spatially
flat FRW universe with a cosmological constant $\Lambda$ and perfect
fluid. The Einstein equations are
%%%
\begin{subequations}
\begin{align}
 & \dot\al^2=\frac{\Lambda}{3}+\frac{\rho_1}{3}, \label{eq:frweq1}\\
 & \ddot\al+\frac{3}{2}\dot\al^2=-\frac{p_1}{2}, 
\end{align}
\end{subequations}
%%%
where $\al$ is a logarithm of a scale factor of the universe, $\rho_1$
and $p_1$ are the energy density and the pressure of perfect fluid,
respectively. The equation of  state is assumed to be
$p_1=(\Gamma-1)\rho_1$ where $\Gamma$ is a constant. Conservation
equation for fluid is
%%%
\begin{equation}
 \dot\rho_1+3\dot\al(\rho_1+p_1)=0,
\end{equation}
%%%
and the solution is
%%%
\begin{equation}
 \rho_1=\frac{c_1}{e^{3\Gamma\al}},
\end{equation}
%%%
where $c_1$ is a constant of integration. By substituting this solution to
Eq.(\ref{eq:frweq1}), we obtain
%%%
\begin{equation}
 \label{eq:orig}
 \dot\al^2=\frac{\Lambda}{3}+\frac{c_1}{e^{3\Gamma\al}}.
\end{equation}
%%%
We solve this equation perturbatively by assuming the second term of
the right-hand side is small:
%%%
\begin{equation}
 \al=\al_0+\al_1+\cdots.
\end{equation}
%%%
The solution of the scale factor up to the first-order of perturbation
is given by
%%%
\begin{equation}
\label{eq:frwnaive}
 a(t)=a_0 \,e^{H_0\,t}\left[1-\frac{c_1\,a_0^{-3\Gamma}}{2\Gamma\Lambda}\left(e^{-3\Gamma H_0 t}-e^{-3\Gamma H_0t_0}\right) \right],
\end{equation}
%%%
where $H_0=\sqrt{\Lambda/3}$, $a_0$ and $t_0$ are constants of
integration of the zeroth-order and the first-order, respectively. We
regard the first-order solution is  secular  and apply the
renormalization-group method\cite{chen,kunihiro,nozaki,nambu}.

We redefine the zeroth-order integration constant $a_0$ as
%%%
\begin{equation}
 a_0=a_0^R(\mu)+\del a_0(t_0;\mu),
\end{equation}
%%%
where $\mu$ is a renormalization point and $\del a_0$ is a counterterm
which absorbs the secular term $\propto e^{-3\Gamma H_0t}$ which
diverges as $t\rightarrow 0$.  The naive solution (\ref{eq:frwnaive})
can be written
%%%
\begin{align}
 a(t)&=e^{H_0\,t}\left[a_0^R(\mu)+\del a_0(t_0;\mu) -\frac{c_1\,(a_0^R)^{1-3\Gamma}}{2\Gamma\Lambda}\left(e^{-3\Gamma H_0 t}-e^{-3\Gamma H_0 \mu}+e^{-3\Gamma H_0 \mu}-e^{-3\Gamma H_0t_0}\right) \right] \notag \\
 &=e^{H_0\,t}\left[a_0^R(\mu) -\frac{c_1\,(a_0^R)^{1-3\Gamma}}{2\Gamma\Lambda}\left(e^{-3\Gamma H_0 t}-e^{-3\Gamma H_0 \mu}\right) \right], \label{eq:naive2}
\end{align}
%%%
where we have chosen the counterterm $\del a_0$ so as to absorb the
$(e^{-3\Gamma H_0 \mu}-e^{-3\Gamma H_0t_0})$-dependent term:
%%%
\begin{equation}
 \label{eq:frwcounter}
 \del a_0(t_0;\mu)=a_0^R(t_0)-a_0^R(\mu)=\frac{c_1\,(a_0^R)^{1-3\Gamma}}{2\Gamma\Lambda}\left(e^{-3\Gamma H_0 \mu}-e^{-3\Gamma H_0 t_0}\right).
\end{equation}
%%%
This defines the renormalization transformation
%%%
\begin{equation}
 {\mathcal R}_{\mu,t_0}~:~a_0^R(t_0)\longmapsto a_0^R(\mu)=a_0^R(t_0)-\frac{c_1\,(a_0^R)^{1-3\Gamma}}{2\Gamma\Lambda}\left(e^{-3\Gamma H_0 \mu}-e^{-3\Gamma H_0 t_0}\right),
\end{equation}
%%%
and this transformation forms the Lie group up to the first-order of
the perturbation. We can obtain $a_0(\mu)$ for arbitrary large value
of $(e^{-3\Gamma H_0 \mu}-e^{-3\Gamma H_0t_0})$ by assuming the
property of the Lie group and this makes it possible to produce a
globally uniform approximated solution of the original differential
equation \eqref{eq:orig}. The renormalization group equation is obtained by
differentiating Eq. \eqref{eq:frwcounter} with respect to $\mu$ and
setting $t_0=\mu$:
%%%
\begin{equation}
 \frac{\pa}{\pa\mu}a_0^R(\mu)=-\frac{c_1}{2\Gamma\Lambda}(a_0^R)^{1-3\Gamma}\frac{\pa}{\pa\mu}e^{-3\Gamma H_0\mu}.
\end{equation}
%%%
The renormalized solution is obtained by equating $\mu=t$ in (\ref{eq:naive2}):
%%%
\begin{equation}
\label{eq:renosol}
 a^R(t)=e^{H_0t}\left(\text{const}-\frac{3c_1}{2\Lambda}e^{-3\Gamma H_0 t}\right)^{1/(3\Gamma)}.
\end{equation}
%%%%%%%%%%%%%%%%%%%%%%%%%%%%%%%%%%%%%%%%%%%%%%%%%%%%%%%%%%%%%%%%%%%%
\section{renormalization-group approach to the cosmological back reaction problem}

We treat the cosmological back reaction problem using a perturbation
approach. Let us assume that the metric is expanded as follows:
%%%
\begin{equation}
 g_{ab}=\overset{(0)}{g}_{ab}+\overset{(1)}{g}_{ab}+\overset{(2)}{g}_{ab}+\cdots.
\end{equation}
%%%
 $\overset{(0)}{g}_{ab}$ is the background FRW metric and represents
a homogeneous and isotropic space. $\overset{(1)}{g}_{ab}$ is the
metric of the first-order linear perturbation. We assume that the
spatial average of the first-order perturbation vanishes:
%%%
\begin{equation}
 \langle\overset{(1)}{g}_{ab}\rangle=0,
\end{equation}
%%%
where $\langle\cdots\rangle$ means the spatial average with respect to
the background FRW metric. $\overset{(2)}{g}_{ab}$ is the second-order
metric and contains nonlinear effect caused by the first-order linear
perturbation. This nonlinearity produces homogeneous and isotropic
zero mode part of the second-order metric. That is, 
%%%
\begin{equation}
  \langle\overset{(2)}{g}_{ab}\rangle\neq0.
\end{equation}
%%%
As we want to interpret the zero mode part of the metric as the
background FRW metric, we must redefine the background metric as
follows:
%%%
\begin{equation}
 \overset{(0)}{g}_{ab}\longrightarrow\overset{(0)}{g}_{ab}+\langle\overset{(2)}{g}_{ab}\rangle. \label{eq:naive-back}
\end{equation}
%%%
This is the back reaction caused by the nonlinearities of the
fluctuation and it changes the background metric. But in general, the
second-order perturbation term will dominate the background metric in
course of time and this simple prescription does not work to get the
long time behaviour of the system.  Furthermore, the meaning of the
gauge invariance is not obvious in the second-order quantity and we
cannot adopt Eq.\eqref{eq:naive-back} as the definition of the
background metric because the gauge transformation changes the
definition of the background metric. By using the renormalization
group method with the second-order gauge invariant quantities, we
resolve these problems and can obtain the effective scale factor of
the FRW universe with the back reaction in a gauge-invariant
manner\cite{nambu2}.
 
We consider a spatially flat FRW universe with scalar type first-order
perturbation. As a matter field, we consider a minimally coupled
scalar field $\chi$. The metric and the scalar field are
%%%
\begin{equation}
\begin{split}
 ds^2&=-(1+2\phi+2\phi_2)dt^2+a^2(t)[(1-2\psi-2\psi_2)\del_{ij}+2E_{,ij}]dx^idx^j, \\
 \chi&=\chi_0(t)+\chi_1+\chi_2,
\end{split}
\end{equation}
%%%
where $a(t)$ is a scale factor of the background FRW universe,
$\chi_0(t)$ is the background scalar field, $\phi,\psi,E,\chi_1$ are
the first-order variables and $\phi_2,\psi_2,\chi_2$ are the
second-order zero mode variables. To obtain the back reaction on the
FRW universe, it is sufficient to consider only the zero mode part of
the second-order perturbation. We use a comoving gauge in which the
fluctuation of the scalar field vanishes :$\chi_1=\chi_2=0$. The
second-order gauge invariant quantity which is invariant under the
first-order gauge transformation is given by\cite{nambu2}
%%%
\begin{equation}
\begin{split}
 &\langle\overset{(2)}{Q}_{ab}\rangle=\langle\overset{(2)}{g}_{ab}\rangle+\langle{\mathcal L}_{\boldsymbol{X}}\overset{(1)}{g}_{ab}\rangle+\frac{1}{2}\langle{\mathcal L}_{\boldsymbol{X}}^2\overset{(0)}{g}_{ab}\rangle, \\
 &~X^\mu=(0, -\del^{ij}E_{,j}).
\end{split}
\end{equation}
%%%
and its components are
%%%
\begin{subequations}
\begin{align}
 & \langle\overset{(2)}{Q}_{00}\rangle= -2\phi_2+a^2\sum_{\bk}k^2\dot E_{\bk}\dot E^{*}_{\bk} ,\\
 & \langle\overset{(2)}{Q}_{ij}\rangle=a^2\left(-2\psi_2+\sum_{\bk}\left(\frac{2k^2}{3}\psi_{\bk}E_{\bk}^{*}-\frac{k^4}{3}E_{\bk}E_{\bk}^{*} \right)\right)\del_{ij}.
\end{align}
\end{subequations}
%%%
where $\psi_{\bk}, E_{\bk}$ are Fourier component of $\psi(t,\bx), E(t,\bx)$:
%%%
\begin{equation}
  \begin{split}
   &\psi(t,\bx)=\sum_{\bk}e^{i\,\bk\cdot\bx}\,\psi_{\bk}(t),\quad \psi_{\bk}^{*}=\psi_{-\bk}, \\
   &E(t,\bx)=\sum_{\bk}e^{i\,\bk\cdot\bx}\,E_{\bk}(t),\quad E_{\bk}^{*}=E_{-\bk}.
  \end{split}
\end{equation}
%%%
The line element of the FRW universe obtained from  the second-order gauge
invariant variables is
%%%
\begin{multline}
 \langle ds^2\rangle=-\left(1+2\phi_2-a^2\sum_{\bk}k^2\dot E_{\bk}\dot E^{*}_{\bk}\right)dt^2\\
 +a^2(t)\left(1-2\psi_2+\sum_{\bk}\left(\frac{2k^2}{3}\psi_{\bk}E_{\bk}^{*}-\frac{k^4}{3}E_{\bk}E_{\bk}^{*}\right) \right)d\bx^2.
\end{multline}
%%%
By the second-order coordinate transformation of time
%%%
\begin{equation}
 t=T-\int dT\left(\phi_2-\frac{a^2}{2}\sum_{\bk}k^2\dot E_{\bk}\dot E_{\bk}^{*}\right),
\end{equation}
%%%
we obtains the metric of the FRW universe in a synchronous form:
%%%
\begin{equation}
\label{eq:frwmetric}
\begin{split}
 &\langle ds^2\rangle=-dT^2+(a^2+\del a^2)d\bx^2,   \\
 &\del
 a^2=a^2(T)\left(\sum_{\bk}\left(\frac{2k^2}{3}\psi_{\bk}E_{\bk}^{*}-\frac{k^4}{3} E_{\bk}E_{\bk}^{*}\right) -2\psi_2-2H\int dT\left(\phi_2-\frac{a^2}{2}\sum_{\bk}k^2\dot E_{\bk}\dot E_{\bk}^{*}\right)\right). 
\end{split}
\end{equation}
%%%
The second-order term $\del a^2$ represents the back reaction of
inhomogeneity on the FRW universe. In the context of perturbative
expansion, the effective scale factor of the FRW universe with the
back reaction can be written as
%%%
\begin{equation}
  \label{eq:naivesc}
  a_{\text{eff}}=a\left(1+\frac{\del a^2}{2a^2}\right).
\end{equation}
%%%
However in general, the second-order correction $\del a^2$ may
dominate the background scale factor $a^2$ as time goes on, and the
prescription of perturbation will break down.  We can go beyond the
perturbation by using the renormalization-group method. We absorb
$\del a^2$ to a constant of integration of the background scale factor
and this procedure yields the renormalization-group equation. By
solving the renormalization group equation, we obtain a renormalized
scale factor which gives more accurate late time behaviour of the
system than the naive perturbative expression (\ref{eq:naivesc}).
%%%%%%%%%%%%%%%%%%%%%%%%%%%%%%%%%%%%%%%%%%%%%%%%%%%%%%%%%%%5
\section{the back reaction in the inflationary universe} 

In this section, we investigate the back reaction problem in the
inflationary universe using the renormalization-group method. We work
in the comoving gauge $\chi_1=\chi_2=0$. 
The background equations are
%%%
\begin{subequations}
\begin{align}
  & 3H^2=\frac{1}{2}\dot\chi_0^2+V(\chi_0), \\
  & 2\dot H+3H^2=-\frac{1}{2}\dot\chi_0^2+V(\chi_0), \\
  &  \ddot\chi_0+3H\dot\chi_0+V'(\chi_0)=0.
\end{align}
\end{subequations}
%%%

The first-order equations  are
%%%
\begin{subequations}
\begin{align}
&  3H(\dot\psi_{\bk}+H\phi_{\bk})+\frac{k^2}{a^2}\psi_{\bk}+k^2 H\dot E_{\bk}=\frac{\dot\chi_0^2}{2}\phi_{\bk}, 
  \label{eq:1a}\\
&  \dot\psi_{\bk}+H\phi_{\bk}=0, \label{eq:1b}\\
&  \ddot\psi_{\bk}+H(\dot\phi_{\bk}+3\dot\psi_{\bk})+(3H^2+2\dot H)\phi_{\bk}=-\frac{\dot\chi_0^2}{2}\phi_{\bk}, 
  \label{eq:1c} \\
&  \ddot E_{\bk}+3H\dot E_{\bk}+\frac{1}{a^2}(\psi_{\bk}-\phi_{\bk})=0, \label{eq:1d}\\
&  2V'(\chi_0)\phi_{\bk}-\dot\chi_0\left(\dot\phi_{\bk}+3\dot\psi_{\bk}+k^2\dot E_{\bk}\right)=0. \label{eq:1e}
\end{align}
\end{subequations}
%%%
Equations \eqref{eq:1a},\eqref{eq:1b},\eqref{eq:1c} and \eqref{eq:1c}
are the first-order Einstein equations, and Eq.\eqref{eq:1e} is the
first-order equation of motion of the scalar field.  The spatial
curvature perturbation $\psi$ obeys the following single equation:
%%%
\begin{equation}
 \ddot\psi_{\bk}+\left(3H-\frac{2\dot H}{H}+\frac{2\ddot\chi_0}{\dot\chi_0}\right)\dot\psi_{\bk}+\frac{k^2}{a^2}\psi_{\bk}=0.  \label{eq:psi}
\end{equation}
%%%

The second order equations for the zero mode variables $\phi_2, \psi_2$  are
%%%
\begin{subequations}
\begin{align}
  &6H\dot\psi_2+2(3H^2+\dot H)\phi_2 = \notag \\ 
  &\qquad\qquad \sum_{\bk}\Bigl[-\frac{2k^4}{a^2}E_{\bk}\psi^{*}_{\bk}-\frac{5k^2}{a^2}\psi_{\bk}\psi^{*}_{\bk}+2k^2\dot E_{\bk}\dot\psi^{*}_{\bk} +
  {3}\dot\psi_{\bk}\dot\psi^{*}_{\bk} -12H(\psi_{\bk}\dot\psi^{*}_{\bk}-\phi_{\bk}\dot\psi^{*}_{\bk})
  \notag \\
  &\qquad\qquad\qquad +4(3H^2+\dot H)\phi_{\bk}\phi^{*}_{\bk}
    -4H(k^4 E_{\bk}\dot E^{*}_{\bk}-k^2\phi_{\bk}\dot E^{*}_{\bk}+k^2\psi_{\bk}\dot E^{*}_{\bk}+k^2 E_{\bk}\dot\psi^{*}_{\bk})\Bigr],
    \label{eq:2a}
\end{align}
%%%
\begin{align}
  &2(\dot H+3H^2)\phi_2+2H(\dot\phi_2+3\dot\psi_2)+2\ddot\psi_2 =
  \notag \\
  &\qquad \sum_{\bk}\Bigl[-\frac{2k^2}{3a^2}\left(k^2E_{\bk}(\psi^{*}_{\bk}-\phi^{*}_{\bk})+\phi_{\bk}\phi^{*}_{\bk}-3\phi_{\bk}\psi^{*}_{\bk}+\frac{5}{2}\psi_{\bk}\psi^{*}_{\bk}
    \right)-\frac{2k^2}{3}\dot E_{\bk} (k^2\dot E^{*}_{\bk}+\dot\psi^{*}_{\bk}-\dot\phi^{*}_{\bk})
     \notag \\
  &\qquad\qquad +2\dot\psi_{\bk}\dot\phi^{*}_{\bk}-\dot\psi_{\bk}\dot\psi^{*}_{\bk}+4(2H^2+\dot H)\phi_{\bk}\phi^{*}_{\bk}-\frac{4k^2}{3}\ddot E_{\bk}(k^2E^{*}_{\bk}+\psi^{*}_{\bk}-\phi^{*}_{\bk}) \notag \\
  &\qquad\qquad
  -4\ddot\psi_{\bk}\left(\frac{k^2}{3}E^{*}_{\bk}+\psi^{*}_{\bk}-\phi^{*}_{\bk}\right) 
   -4H(-H\phi_{\bk}\phi^{*}_{\bk}-2\phi_{\bk}\dot\phi^{*}_{\bk}-3\phi_{\bk}\dot\psi^{*}_{\bk}+3\psi_{\bk}\dot\psi^{*}_{\bk}) \notag \\
   &\qquad\qquad
   -4k^2 H(k^2 E_{\bk}\dot E^{*}_{\bk} -\phi_{\bk}\dot E^{*}_{\bk}+\psi_{\bk}\dot E^{*}_{\bk}+E_{\bk}\dot\psi^{*}_{\bk})\Bigr],
  \label{eq:2b}
\end{align}
%%%%%%%
\begin{align}
  &-\dot\chi_0(\dot\phi_2+3\dot\psi_2)-2(\ddot\chi_0+3H\dot\chi_0)\phi_2= \notag \\
  &\qquad -\dot\chi_0\sum_{\bk}\Bigl[-\frac{8k^4}{3}E_{\bk}\dot E^{*}_{\bk}+{k^2}(3\dot
  E_{\bk}\phi^{*}_{\bk}+E_{\bk}\dot\phi^{*}_{\bk}) 
   -4k^2(\dot
  E_{\bk}\psi^{*}_{\bk}+E_{\bk}\dot\psi^{*}_{\bk}) \notag \\
  &\qquad\qquad\qquad +4\phi_{\bk}\dot\phi^{*}_{\bk}+{3}\dot\phi_{\bk}\psi^{*}_{\bk}
   +{9}\phi_{\bk}\dot\psi^{*}_{\bk}-12\psi_{\bk}\dot\psi^{*}_{\bk}
  +4\left(3H+\frac{\ddot\chi_0}{\dot\chi_0}\right)\phi_{\bk}\phi^{*}_{\bk}\Bigr].
  \label{eq:2c}
\end{align}
\end{subequations}
%%%
Eqs.\eqref{eq:2a} and \eqref{eq:2b} are the second-order Einstein
equations and Eq.\eqref{eq:2c} is the second-order equation of motion
of the scalar field.

We solve these equations under the condition of slow-rolling inflation
%%%
\begin{equation}
  \left|\frac{\dot H}{H^2}\right|\ll 1,\quad
  \left|\frac{\ddot\chi_0}{H\dot\chi_0}\right|\ll 1.
\end{equation}
%%%
Under this condition,
%%%
\begin{equation}
 H^2\approx \frac{1}{3}V(\chi_0),\quad \dot\chi_0\approx-\frac{V'(\chi_0)}{3H}.
\end{equation}
%%%%%%%%%%%%%%%%%%%%%%%%%%%%%%%%%%%%%%%%%%%%%%%%%
\subsection{Long wavelength mode}

For the long wavelength mode of which wavelength is larger than the
horizon scale $1/H$, the growing mode solution of Eq.(\ref{eq:psi}) is
given by
%%%
\begin{equation}
 \psi_{\bk}=\psi_0(k)\left[1+\frac{k^2}{2a^2H^2}\right]+O(k^4),
\end{equation}
%%%
where $\psi_0(k)$ is an arbitrary function of $k$ and satisfies
$\psi_0(k)=0$ as $k\rightarrow 0$ to ensure $\langle\psi\rangle=0$. If
we take only $O(k^0)$ term $\psi_0(k)$, the right-hand sides of the
second-order equations become zero. We must take account of $O(k^2)$
term to get the back reaction effect. By using this as the first order
solution, other first-order quantities are given by
%%%
\begin{equation}
\begin{split}
 &  \phi\approx\psi_0\frac{k^2}{a^2H^2},\quad E\approx\frac{\psi_0}{2}\frac{1}{a^2H^2}, \\
 & \dot\psi\approx-\psi_0\frac{k^2}{a^2H},\quad \dot\phi\approx -2\psi_0\frac{k^2}{a^2H},\quad \dot E\approx -\psi_0\frac{1}{a^2H}
\end{split}
\end{equation}
%%%
Then, the second-order equations become
%%%
\begin{subequations}
\begin{align}
  & \dot\psi_2+H\phi_2\approx \frac{11}{6a^2H}\sum_{\bk}k^2|\psi_0|^2, \\
  & \ddot\psi_2+H\dot\phi_2+3H(\dot\psi_2+H\phi_2)
    \approx \frac{11}{6a^2}\sum_{\bk}k^2|\psi_0|^2, \\
  &  6H\phi_2+\dot\phi_2+3\dot\psi_2\approx \frac{10}{a^2H}\sum_{\bk}k^2|\psi_0|^2,
\end{align}
\end{subequations}
%%%
and the second-order zero mode solution is
%%%
\begin{equation}
 \psi_2\approx \frac{4}{3a^2H^2}\sum_{\bk}k^2|\psi_0|^2,\quad 
 \phi_2\approx \frac{9}{2a^2H^2}\sum_{\bk}k^2|\psi_0|^2
\end{equation}
%%%
The metric of FRW universe (\ref{eq:frwmetric}) is given by
%%%
\begin{equation}
 \label{eq:frwlong}
 \langle ds^2\rangle=-dT^2+a^2(T)\left(1+\frac{7}{3a^2H^2}\sum_{\bk}k^2|\psi_0|^2
 \right)d\boldsymbol{x}^2.
\end{equation}
%%%

At this stage, we apply the renormalization-group method to obtain the
effective scale factor which includes the back reaction effect. The
background scale factor can be written
%%%
\begin{equation}
 a(T)=a_0\,\tilde a(T),
\end{equation}
%%%
where $a_0$ is a constant of integration which reflects the freedom of
rescaling of the scale factor. We redefine the constant $a_0$ so as to
absorb the second order correction of the scale factor. The
renormalization-group equation is given by
%%%
\begin{equation}
 \frac{\pa a_0^2}{\pa(1/\tilde a^2)}=\frac{7}{3H^2}\sum_{\bk}k^2|\psi_0|^2,
\end{equation}
%%%
and the solution is
%%%
\begin{equation}
 a_0(T)=\left(\text{const}+\frac{7}{3\tilde a^2H^2}\sum_{\bk}k^2|\psi_0|^2\right)^{1/2}.
\end{equation}
%%%
The renormalized metric is
%%%
\begin{equation}
\begin{split}
 &\langle ds^2\rangle=-dT^2+\left(a^R(T)\right)^2d\bx^2, \\
 &a^R(T)=\tilde a(T)\left(\text{const}+\frac{7}{3\tilde a^2H^2}\sum_{\bk}k^2|\psi_0|^2\right)^{1/2}.
\end{split}
\end{equation}
%%%
Comparing with the analysis of the Sec. II, the renormalized scale
factor is the same as that of the FRW universe with
$\Gamma=\frac{2}{3}, c_1<0$. We conclude that the back reaction of the
long wavelength scalar perturbation on the FRW universe is equivalent
to a positive spatial curvature. But because of the inflationary
expansion of the universe, the back reaction effect decays as
$e^{-2Ht}$ and becomes negligible.

%%%%%%%%%%%%%%%%%%%%%%%%%%%%%%%%%%%%%%%%%%%%%%
\subsection{Short wavelength mode}

For the short wavelength mode of which wavelength is smaller than the
horizon scale $1/H$, the first-order solution of $\psi$ is given by
the WKB form
%%%
\begin{equation}
 \psi\approx\frac{\psi_0}{a}\cos(k\eta),
\end{equation}
%%%
where $\displaystyle{\psi_0\equiv\frac{H}{\dot\chi_0}C(k)}$, $C(k)$ is
an arbitrary function of $k$ and
$\displaystyle{\eta=\int\frac{dt}{a}}$. In the inflationary universe,
$\psi_0$ becomes approximately constant in time. Other first-order
quantities are
%%%
\begin{equation}
 \dot E\approx-\frac{1}{Ha^2}\psi,\quad  k^2 E\approx 3\psi+\frac{\dot\psi}{H}.
\end{equation}
%%%%
Using these solution, the second-order equation  becomes
%%%
\begin{subequations}
\begin{align}
  & \dot\psi_2+H\phi_2\approx 0, \\
  & \ddot\psi_2+H\dot\phi_2+3H(\dot\psi_2+H\phi_2)\approx 0, \\
  & \dot\phi_2+3\dot\psi_2+6H\phi_2\approx\frac{8}{3a^4H}\sum_{\bk}k^2|\psi_0|^2,
\end{align}
\end{subequations}
%%%
where we have omitted oscillatory terms in the right-hand sides
because they does not contribute to the secular behavior of the zero
mode solution. The second-order zero mode metric is given by
%%%
\begin{equation}
  \phi_2\approx -\frac{8}{3H^2a^4}\sum_{\bk}k^2|\psi_0|^2,\quad 
  \psi_2\approx -\frac{2}{3H^2a^4}\sum_{\bk}k^2|\psi_0|^2.
\end{equation}
Therefore the metric of the FRW universe (\ref{eq:frwmetric}) becomes
%%%
\begin{equation}
\label{eq:frwshort}
 ds^2=-dT^2+a^2(T)\left(1-\frac{7}{24a^4H^2}\sum_{\bk}k^2|\psi_0|^2 \right)d\boldsymbol{x}^2.
\end{equation}
%%%

The renormalization group equation becomes
%%%
\begin{equation}
 \frac{\pa a_0^2}{\pa(1/\tilde a^4)}=-\frac{7}{24a_0^2H^2}\sum_{\bk}k^2|\psi_0|^2,
\end{equation}
%%%
and the solution is
%%%
\begin{equation}
 a_0(T)=\left(\text{const}-\frac{7}{24\tilde a^4H^2}\sum_{\bk}k^2|\psi_0|^2 \right)^{1/4}.
\end{equation}
%%%
The renormalized metric is
%%%
\begin{equation}
\begin{split}
 &\langle ds^2\rangle=-dT^2+\left(a^R(T)\right)^2d\bx^2, \\
 &a^R(T)=\tilde a(T)\left(\text{const}-\frac{7}{24\tilde a^4H^2}\sum_{\bk}k^2|\psi_0|^2 \right)^{1/4}.
\end{split}
\end{equation}
%%%
Comparing with the result of Sec. II, the renormalized scale factor is
the same as that of the FRW universe with $\Gamma=\frac{4}{3}, c_1>0$.
The effect of the back reaction of the short wavelength scalar
perturbation is same as a radiation fluid.
%%%%%%%%%%%%%%%%%%%%%%%%%%%%%%%%%%%%%%%
\subsection{Long wavelength quantum fluctuation}

We consider the back reaction due to the quantum fluctuation in the
inflationary universe. By quantizing the first order perturbation, the
normalization of the mode function is determined. For the quantum
fluctuation, the operation of spatial average $\langle\cdots\rangle$
must be replaced with an expectation value with a suitable vacuum
state. Introducing a variable
$\displaystyle{\del\chi\equiv\frac{\dot\chi_0}{H}\psi}$,
Eq.(\ref{eq:psi}) becomes\cite{hwang}
%%%
\begin{equation}
\label{eq:muk}
\del\ddot\chi_{\bk}+3H\del\dot\chi_{\bk}+\left(\frac{k^2}{a^2}+V''+\frac{2\dot H}{H}\left(3H-\frac{\dot H}{H}+\frac{2\ddot\chi_0}{\dot\chi_0}\right) \right)\del\chi_{\bk}=0.
\end{equation}
%%%
We quantize the variable $\del\chi$:
%%%
\begin{subequations}
\begin{align}
 & \del\hat\chi(\bx,t)=\int\frac{d^3k}{(2\pi)^{3/2}}\left[\hat a_{\bk}\del\chi_{\bk}(t)e^{i\bk\cdot\bx}+\hat a_{\bk}^{\dagger}\del\chi_{\bk}^{*}(t)e^{-i\bk\cdot\bx}\right], \\
 & \qquad [\hat a_{\bk},\hat a_{\bk'}]=[\hat a_{\bk}^{\dagger},\hat a_{\bk'}^{\dagger}]=0,\quad [\hat a_{\bk},\hat a_{\bk'}^{\dagger}]=\hbar\, \del^3(\bk-\bk'), \\
 &\qquad \del\chi_{\bk}\,\del\dot\chi_{\bk}^{*}-\del\chi_{\bk}^{*}\,\del\dot\chi_{\bk}=\frac{i}{a^3}.
\end{align}
\end{subequations}
%%%
Under the slow-rolling condition, Eq.(\ref{eq:muk}) is approximately
same as the mode equation of the scalar field on a fixed deSitter
background, and the solution of the mode function is given by
%%%
\begin{equation}
\begin{split}
 & \del\chi_{\bk}(t)\approx\frac{\sqrt{\pi}}{2}H\eta^{3/2}\left[c_1(k)H_\nu^{(1)}(k\eta)+c_2(k)H_\nu^{(2)}(k\eta)\right], \\
 & \quad\eta\approx-\frac{1}{aH},\quad\nu\approx\frac{3}{2}-\frac{V''(\chi_0)}{3H^2},\quad |c_2(k)|^2-|c_1(k)|^2=1,
\end{split}
\end{equation}
%%%
where $H_\nu^{(1,2)}$ is a Hankel function.  We choose Bunch-Davies
vacuum $c_2=1, c_1=0$. The power spectrum for $\del\chi$ in the long
wavelength is given by
%%%
\begin{equation}
 P_{\del\chi}(k,t)\equiv\hbar\frac{k^3}{2\pi^2}|\del\chi_{\bk}(t)|^2\approx\hbar\left(\frac{H}{2\pi}\right)^2\left(\frac{k}{aH}\right)^{3-2\nu},
\end{equation}
%%%
and
%%%
\begin{equation}
\begin{split}
& \langle\del\chi^2\rangle=\hbar\sum_{\bk}|\del\chi_{\bk}|^2=\int_H^{aH}d\ln k\,P_{\del\chi}(k,t)\approx\frac{3\hbar H^4}{8\pi^2V''(\chi_0)^2}\left[1-\exp\left(-\frac{2V''(\chi_0)^2}{3H}t\right)\right], \\
& \langle(\nabla\del\chi)^2 \rangle=\hbar\sum_{\bk}k^2|\del\chi_{\bk}(t)|^2=\int_H^{aH}d\ln k\,k^2P_{\del\chi}(k,t)\approx \frac{\hbar}{2}\left(\frac{H}{2\pi}\right)^2H^2(a^2-1),
\end{split}
\end{equation}
%%%
where we cut off the infrared and ultraviolet contributions, which is
conventionally used regularization.  Therefore
%%%
\begin{equation}
 \langle(\nabla\psi)^2\rangle=\hbar\left(\frac{H}{\dot\chi_0}\right)^2\sum_{\bk}k^2|\del\chi_{\bk}|^2\approx\frac{\hbar}{2}\left(\frac{H}{\dot\chi_0}\right)^2\left(\frac{H}{2\pi}\right)^2H^2(a^2-1),
\end{equation}
%%%
and the metric of FRW universe (\ref{eq:frwlong}) becomes
%%%
\begin{align}
 \langle ds^2\rangle&=-dT^2+a^2(T)\left(1+\frac{7}{3a^2H^2}\langle(\nabla\psi)^2\rangle \right)d\bx^2 \notag \\
 &=-dT^2+a^2(T)\left[1+\frac{7\hbar}{6}\left(\frac{H}{\dot\chi_0}\right)^2\left(\frac{H}{2\pi}\right)^2\left(1-\frac{1}{a^2}\right)\right]d\bx^2.\label{eq:frwquantum}
\end{align}
%%%

The renormalization-group equation is
%%%
\begin{equation}
 \frac{\pa a_0^2}{\pa(1/\tilde a^2)}=-\frac{7\hbar}{6}\left(\frac{H}{\dot\chi_0}\right)^2\left(\frac{H}{2\pi}\right)^2,
\end{equation}
%%%
and the renormalized metric is
%%%
\begin{equation}
\begin{split}
 & \langle ds^2\rangle=-dT^2+\left(a^R(T)\right)^2 d\bx^2, \\
 & a^R(T)=\tilde a(T)\left(\text{const}-\frac{7\hbar}{6\tilde a^2}\left(\frac{H}{\dot\chi_0}\right)^2\left(\frac{H}{2\pi}\right)^2\right)^{1/2}.
\end{split}
\end{equation}
%%%
Comparing with the result of Sec. II, the renormalized scale factor is 
the same as that of the FRW universe with $\Gamma=\frac{2}{3}, c_1<0$. 
The effect of the back reaction of the
quantum fluctuation is equivalent to a negative spatial curvature.

%%%%%%%%%%%%%%%%%%%%%%%%%%%%%%%%%%%%%%%%%%%%%%%%%%%%%%%%%%%%%
\section{summary and discussion}

We have investigated the back reaction problem in the inflationary
universe using the renormalization-group method. By renormalizing the
second-order gauge-invariant zero mode metric which appears by the
nonlinear effect of Einstein's equation, we obtained the effective
scale factor which include the back reaction of the scalar-type
fluctuation. For the long wavelength classical perturbation, the
effect of the back reaction is equivalent to a positive spatial
curvature. For the short wavelength perturbation, the back reaction is
equivalent to a radiation fluid. For the long wavelength quantum
fluctuation, the effect of the back reaction is same as a negative
spatial curvature. In any case, the effect of the back reaction
quickly decays and becomes negligible, and does not alter the
expansion of the inflationary universe.

Our result on the back reaction of the long wavelength perturbation is
different from the analysis of Abramo and
co-workers\cite{mukuhanov1,abramo1,abramo2,abramo3}. They used the
longitudinal gauge and obtained the result that the back reaction due
to the long wavelength perturbation works as a negative cosmological
constant.  To understand why this discrepancy occurs, we have performed the
calculation of the back reaction using the longitudinal gauge(see
Appendix). The point is the form of the first order solution which is used to 
evaluate the back reaction effect.  In the long wavelength limit, the growing 
mode solution of $\psi$ in the longitudinal gauge is given by
%%%
\begin{equation}
 \psi=\frac{H}{a}\int_{t_0}^t dt\,a\frac{\dot H}{H^2}\approx
 -\left(\frac{a}{H}\right)_{t=t_0}\frac{H}{a}+\frac{\dot H}{H^2}+\cdots.
\end{equation}
%%%
They used $\frac{\dot H}{H^2}$ as the first order solution. But the
obtained second order solution is canceled to be zero by choosing an
appropriate homogeneous solution of the second order equation. Hence
we should take $\frac{H}{a}$ as the first order solution to observe
the back reaction effect.  Our calculation in the longitudinal gauge
shows that we have the same back reaction effect as the comoving gauge
case.

The back reaction becomes important in the preheating stage of the
universe. For a massive scalar field, the scalar field oscillates
around the minimum of the potential and the scale factor grows as
$a\sim t^{2/3}$. We calculated the back reaction of the long
wavelength perturbation at this stage and the renormalized scale
factor is given by
%%%
\begin{equation}
 a^R(t)=T^{2/3}\left(\text{const}-c\,T^{2/3}\sum_{\bk}k^2|\psi_0(k)|^2 \right)^{1/2},
\end{equation}
%%%
where $c$ is a numerical factor. The back reaction of the long
wavelength scalar perturbation is the same as the effect of a positive
spatial curvature and this is the same as the case of dust dominated
universe\cite{nambu2}. In the preheating stage of the universe, the
back reaction slows down the expansion of the universe and its effect
cannot be negligible.

The renormalization-group method can be viewed as a tool of system reduction. The renormalzation-group equation corresponds to the amplitude equation which describes slow 
 motion dynamics in the original system. We can describe complicated dynamics contained in the original equation by extracting a simpler representation using the renormalization-group equation. For the cosmological back reaction problem, we can reduce the Einstein equation to the FRW equation and this is nothing but a back reaction equation. In this paper, we perturbatively solved the Einstein equation and obtained the solution of the FRW equation with the back reaction effect and we does not derive the back reaction equation. It is possible to derive the back reaction equation by applying the renormalization-group method to the equation of motion directly. We will report this subject in a separate publication.

%%%%%%%%%%%%%%%%%%%%%%%%%%%%%%%%%%%%%%%%%%%%%%%%%%%%%%%%%%%%%%%%%%%%%%%%%
\acknowledgements{ The author would like to thank L. R. Abramo for his
  comment on our previous paper\cite{nambu2} and suggestion on the
  back reaction problem in the inflationary universe. This work was
  supported in part by a Grant-In-Aid for Scientific Research of the
  Ministry of Education, Science, Sports and Culture of Japan
  (11640270).}
%%%%%%%%%%%%%%%%%%%%%%%%%%%%%%%%%%%%%%%%%%%%%%%%%%%%%%%%%%%%%%%%%%%%%%%%%%
%%%%%%%%%%%%%%%%%%%%%%%%%%%%%%%%%%%%%%%%%%%%%%%%%%%%%%%%%%%%%%%%%%%%%%%%%%

\appendix
\section*{back reaction in longitudinal gauge}
In this appendix, we present the calculation of the back reaction
using the longitudinal gauge $\phi=\psi, E=0$ to check the gauge
independence of the back reaction effect. The metric is
\begin{equation}
 ds^2=-(1+2\psi+2\phi_2)dt^2+a^2(1-2\psi-2\psi_2)d\boldsymbol{x}^2.
\end{equation}
The first-order equations are
%%%
\begin{subequations}
\begin{align}
& \qquad 3H(\dot\psi_{\bk}+H\psi_{\bk})+\frac{k^2}{a^2}\psi_{\bk}=\frac{1}{2}(-\dot\chi_0 \dot\chi_1{}_{\bk}-V'(\chi_0)\chi_1{}_{\bk}+\dot\chi_0^2 \psi_{\bk}), \\
& \qquad \dot\psi_{\bk}+H\psi_{\bk}=\frac{1}{2}\dot\chi_0 \chi_1{}_{\bk}, \\
& \qquad \ddot\psi_{\bk}+4H\dot\psi_{\bk}+(2\dot H+3H^2)\psi_{\bk}=-\frac{1}{2}(\dot\chi_0^2 \psi_{\bk}-\dot\chi_0\dot\chi_1{}_{\bk}+V'(\chi_0) \chi_1{}_{\bk}), \\
& \qquad \ddot\chi_1{}_{\bk}+3H\dot\chi_1{}_{\bk}+V''(\chi_0) \chi_1{}_{\bk}+\frac{k^2}{a^2} \chi_1{}_{\bk}+2V'(\chi_0) \psi_{\bk}-4\dot\chi_0 \dot\psi_{\bk}=0.
\end{align}
\end{subequations}
%%%
The evolution equation of $\psi$ becomes
\begin{equation}
 \ddot\psi_{\bk}+\left(H-\frac{2\ddot\chi_0}{\dot\chi_0}\right)\dot\psi_{\bk}+\left(2\dot H-2H\frac{\ddot\chi_0}{\dot\chi_0}+\frac{k^2}{a^2}\right)\psi_{\bk}=0.\label{eq:newton}
\end{equation}
%%%
The second-order equations are
%%%
\begin{subequations}
\begin{multline}
 \label{eq:g00}
 6H\dot\psi_2+2(3H^2+\dot H)\phi_2+\dot\chi_0\dot\chi_2+V'(\chi_0)\chi_2  \\
  =\sum_{\bk}\Bigl[-\frac{5k^2}{a^2}\psi_{\bk}\psi^{*}_{\bk}+4(3H^2+\dot H)\psi_{\bk}\psi^{*}_{\bk}+3\dot\psi_{\bk}\dot\psi^{*}_{\bk} \\
  -\frac{1}{2}\dot\chi_1{}_{\bk}\dot\chi_1^{*}{}_{\bk}-\frac{k^2}{2a^2}\chi_1{}_{\bk}\chi_1^{*}{}_{\bk}+2\dot\chi_0\psi_{\bk}\dot\chi_1^{*}{}_{\bk}-\frac{V''}{2}\chi_1{}_{\bk}\chi_1^{*}{}_{\bk}\Bigr], 
\end{multline}
%%%
 \begin{multline}
 \label{eq:gii}
  2\ddot\psi_2+6H\dot\psi_2+2H\dot\phi_2+2(\dot H+3H^2)\phi_2 +V'(\chi_0)\chi_2-\dot\chi_0\dot\chi_2 \\
  =\sum_{\bk}\Bigl[-\frac{k^2}{3a^2}\psi_{\bk}\psi^{*}_{\bk}+4(3H^2+\dot H)\psi_{\bk}\psi^{*}_{\bk}+8H\psi_{\bk}\dot\psi^{*}_{\bk}+\dot\psi_{\bk}\dot\psi^{*}_{\bk}  \\
  +\frac{1}{2}\dot\chi_1{}_{\bk}\dot\chi_1^{*}{}_{\bk}-\frac{k^2}{6a^2}\chi_1{}_{\bk}\chi_1^{*}{}_{\bk}-2\dot\chi_0\psi_{\bk}\dot\chi_1^{*}{}_{\bk}-\frac{V''}{2}\chi_1{}_{\bk}\chi_1^{*}{}_{\bk}\Bigr],  
 \end{multline}
%%%
\begin{multline} 
\ddot\chi_2+3H\dot\chi_2+V''(\chi_0)\chi_2
-\dot\chi_0\left(6H\phi_2+\dot\phi_2+\frac{2\ddot\chi_0}{\dot\chi_0}\phi_2+3\dot\psi_2 
\right)  \\
  =\sum_{\bk}\Bigl[-\frac{V'''(\chi_0)}{2}\chi_1{}_{\bk}\chi_1^{*}{}_{\bk}-\frac{2k^2}{a^2}\psi_{\bk}\chi_1^{*}{}_{\bk}
+6H\psi_{\bk}\dot\chi_1^{*}{}_{\bk}+4\dot\chi_1{}_{\bk}\dot\psi^{*}_{\bk}
+2\psi_{\bk}\ddot\chi_1^{*}{}_{\bk} \\
-4\dot\chi_0\left[\psi_{\bk}\dot\psi^{*}_{\bk}+\left(\frac{\ddot\chi_0}{\dot\chi_0}+3H
\right)\psi_{\bk}\psi^{*}_{\bk} \right]\Bigr].
\end{multline}
\end{subequations}
%%%
In this gauge, $\phi_2$ and $\psi_2$ are invariant under the
first-order gauge transformation.  The metric of the FRW universe is
given by
\begin{align}
  \langle ds^2\rangle &= -(1+2\phi_2)dt^2+a^2(t)(1-2\psi_2)d\boldsymbol{x}^2 \notag \\
  &=-dT^2+a^2(T)\left(1-2\psi_2-2H\int dt\,\phi_2\right)d\boldsymbol{x}^2,
\end{align}
where
%%%
\begin{equation}
 t=T-\int dT\,\phi_2.
\end{equation}
%%%%%%%%%%%%%%%%%%%%%%%

\subsection{Long wavelength case}

The growing mode solution of Eq.\eqref{eq:newton} in long wavelength
limit is given by
%%%
\begin{equation}
 \psi_{\bk}=D(k)\left[-1+\frac{H}{a}\int_{t_0}^t dt\,a\right] \approx D(k)\frac{\dot H}{H^2}+C(k)\frac{H}{a}.
\end{equation}
%%%
where $C(k), D(k)$ are arbitrary functions of $k$ which satisfy
$C(k)=0, D(k)=0$ as $k\rightarrow 0$ to ensure
$\langle\psi\rangle=0$. Using this solution, the second-order solution
is given by
%%%
\begin{subequations}
\begin{align}
 &\phi_2\approx
 \frac{1}{2}(\ep-3\delta)\ep\sum_{\bk}|D|^2
 +\frac{H}{2a}(\ep+7\delta)\sum_{\bk}(CD^{*}+C^{*}D) -
 \frac{5H^2}{2a^2}\sum_{\bk}|C|^2 ,\\
 &\psi_2\approx\frac{1}{2}(\ep-3\delta)\ep\sum_{\bk}|D|^2
  +\frac{H}{2a}(-3\ep+7\delta)\sum_{\bk}(CD^{*}+C^{*}D)
  -\frac{5H^2}{2a^2}\sum_{\bk}|C|^2, \\
 &\chi_2\approx \frac{1}{2}(\ep+9\delta)\frac{\dot\chi_0}{H}\sum_{\bk}|D|^2+\frac{\dot\chi_0H}{2a^2}\sum_{\bk}|C|^2, 
\end{align}
\end{subequations}
%%%
where $\ep$ and $\delta$ are slow roll parameters
%%%
\begin{equation}
  \ep=\frac{\dot H}{H^2},\quad \del=\frac{\ddot\chi_0}{H\dot\chi_0},
\end{equation}
%%%
and they can be treated as small constant under the considering orders of the
approximation. 

It is important to notice that the second-order equation has the
following homogeneous solution:
%%%
\begin{equation}
 \psi_2^{(\text{homo})}=\phi_2^{(\text{homo})}\approx D_2\frac{\dot H}{H^2}+C_2\frac{H}{a},
\end{equation}
%%%
where $C_2, D_2$ are arbitrary constants. By using the freedom of the
homogenenous solution and the second-order gauge transformation
$t\rightarrow t+B_2\frac{H}{a}$ where $B_2$ is a constant, it is
always possible to reduce the second-order solution as the following form:
%%%
\begin{equation}
 \psi_2=\phi_2\approx -\frac{5H^2}{2a^2}\sum_{\bk}|C|^2,\quad\chi_2\approx \frac{\dot\chi_0}{2}\frac{H}{a^2}\sum_{\bk}|C|^2.
\end{equation}
The metric of the FRW universe is given by
%%%
\begin{equation}
  \langle ds^2\rangle=-dT^2+a^2(T)\left(1+\frac{5H^2}{2a^2}\sum_{\bk}|C|^2\right)d\boldsymbol{x}^2.
\end{equation}
%%%
This expression is same as Eq.\eqref{eq:frwlong} and the effect of the
back reaction is equivalent to a positive spatial curvature. This
result is consistent with the calculation using the comoving gauge.

We can read  the component of the gauge invariant effective energy
momentum tensor from \eqref{eq:g00} and \eqref{eq:gii}. \eqref{eq:g00}
is the time-time component and \eqref{eq:gii} is the space-space
component of the second-order Einstein equation. Hence for the
first-order solution $\phi_1=\psi_1\approx C\frac{H}{a}$,
%%%
\begin{equation}
 \overset{(2)}{\rho}=-\frac{15H^4}{a^2}\sum_{\bk}|C|^2,\quad 
 \overset{(2)}{p}=\frac{5H^4}{a^2}\sum_{\bk}|C|^2,
\end{equation}
%%%
and this gives the equation of state
$\overset{(2)}{p}=-\frac{1}{3}\overset{(2)}{\rho},
\overset{(2)}{\rho}<0$. This corresponds to a positive spatial
curvature and consistent with the analysis using the
renormalization-group method.
%%%%%%%%%%%%%%%%%%%%%%%%%%
\subsection{Short wavelength case}
The first-order solution is
\begin{equation}
 \psi\approx -2C\dot\chi_0\sin(k\eta),\quad \chi_1\approx -C\frac{4k}{a}\cos(k\eta)\quad (\eta\equiv\int\frac{dt}{a})
\end{equation}
%%%
The second-order equations are
%%%
\begin{subequations}
\begin{align}
 & 6H\dot\psi_2+6H^2\phi_2+\dot\chi_0\dot\chi_2+V'\chi_2\approx -8\frac{k^4}{a^4}\sum_{\bk}|C|^2, \\
 & 2\ddot\psi_2+6H\dot\psi_2+2H\dot\phi_2-\dot\chi_0\dot\chi_2+V'\chi_2\approx \frac{8}{3}\frac{k^4}{a^4}\sum_{\bk}|C|^2, \\
 & \ddot\chi_2+3H\dot\chi_2+V''\chi_2-\dot\chi_0(6H\phi_2+\dot\phi_2+3\dot\psi_2)\approx 0.
\end{align}
\end{subequations}
%%%
The second-order solutions are
%%%
\begin{equation}
 \psi_2\approx -\frac{2}{3H^2a^4}\sum_{\bk}k^4|C|^2,\quad
 \phi_2\approx -\frac{4}{H^2a^4}\sum_{\bk}k^4|C|^2,\quad \chi_2\approx 0.
\end{equation}
%%%
The metric of the FRW universe is 
%%%
\begin{equation}
 \langle ds^2\rangle=-dT^2+a^2(T)\left(1-\frac{4}{15H^2a^4}\sum_{\bk}k^4|C|^2\right)d\boldsymbol{x}^2.
\end{equation}
%%%
This result is same as Eq.\eqref{eq:frwshort} and the effect of the
back reaction is equivalent to a radiation. The component of the
effective energy momentum tensor is
%%%
\begin{equation}
 \overset{(2)}{\rho}=\frac{8}{a^4}\sum_{\bk}k^4|C|^2,\quad 
 \overset{(2)}{p}=\frac{8}{3a^4}\sum_{\bk}k^4|C|^2,
\end{equation}
%%%
and the equation of state is that of radiation
$\overset{(2)}{p}=\frac{1}{3}\overset{(2)}{\rho},
\overset{(2)}{\rho}>0$.
%%%%%%%%%%%%%%%%%%%%%%%%%%%%%%%%%%%%%%%%%%%%%%%
%\bibliographystyle{prsty}
%\bibliography{paper}

\end{document}